\documentclass[
reprint,
superscriptaddress,
showpacs,
showkeys,
amsmath, amssymb,
aps,
prl,
]{revtex4-1}

\usepackage{graphicx}% Include figure files
\usepackage{dcolumn}% Align table columns on decimal point
\usepackage{bm}% bold math
\usepackage[colorlinks=true]{hyperref}% add hypertext capabilities
\graphicspath{{figures/}}
\usepackage{todonotes}
\usepackage[separate-uncertainty=false]{siunitx}

\usepackage{xspace}
\newcommand{\rb}{$^{87}$Rb\xspace}
\newcommand{\xyz}{$\ket{x,y,z}$\xspace}
\newcommand{\XYZ}{$\ket{X,Y,Z}$\xspace}

\newcommand{\ket}[1]{\vert#1\rangle}

\newcommand{\reffig}[1]{Fig.~\ref{#1}}
\newcommand{\refeq}[1]{Eq.~\ref{#1}}

\begin{document}

\title{Synthetic clock transitions via continuous dynamical decoupling}
\date{\today}

\author{D. Trypogeorgos}
\email[E-mail: ]{dtrypo@umd.edu}
\author{A. Vald\'es-Curiel}
\affiliation{Joint Quantum Institute, University of Maryland and National
Institute of Standards and Technology, College Park, Maryland, 20742, USA}
\author{N. Lundblad}
\affiliation{Department of Physics and Astronomy, Bates College, Lewiston,
Maine 04240, USA}
\author{I. B. Spielman}
\affiliation{Joint Quantum Institute, University of Maryland and National
Institute of Standards and Technology, College Park, Maryland, 20742, USA}

\begin{abstract}
    Decoherence of quantum systems due to uncontrolled fluctuations of the environment presents fundamental obstacles in quantum science.
    `Clock' transitions which are insensitive to such fluctuations are used to improve coherence, however, they are not present in all systems or for arbitrary system parameters.
    Here, we create a trio of synthetic clock transitions
    using continuous dynamical decoupling in a spin-1 Bose-Einstein condensate in which we observe a reduction of sensitivity to magnetic field noise of up to four orders of magnitude; this work complements the parallel work by Anderson et al.~(submitted, 2017).
    In addition, using a concatenated scheme, we demonstrate suppression of sensitivity to fluctuations in our control fields.
    These field-insensitive states represent an ideal foundation for the next generation of cold atom experiments focused on fragile many-body phases relevant to quantum magnetism, artificial gauge fields, and topological matter.
\end{abstract}

%\pacs{32.80.Qk}

\maketitle
%\tableofcontents

%\begin{widetext}
%long equation goes here
%\end{widetext}

The loss of coherence due to uncontrolled coupling to a fluctuating environment is a limiting performance factor for quantum technologies~\cite{chaudhry_decoherence_2012,myatt_decoherence_2000,schlosshauer_decoherence_2005,viola_dynamical_1998}.
In select cases, first-order insensitive transitions --- `clock' transitions --- can mitigate the deleterious effect of the dominant noise sources, yet in most cases such transitions are absent~\cite{[{Quality oscillators that show remarkably precise and robust dynamics can be found even in biological systems. See for example the repressilator gene, }] potvin-trottier_synchronous_2016}.
Remarkably, under almost all circumstances, clock transitions can be synthesized using dynamical decoupling protocols.
These protocols involve driving the system with an external oscillatory field, resulting in a dynamically protected `dressed' system.
A number of dynamical decoupling protocols, pulsed or continuous, have been shown to isolate quantum systems from low-frequency environmental noise~\cite{cohen_continuous_2017,fanchini_continuously_2007,aharon_fully_2016,biercuk_optimized_2009,cai_robust_2012,bermudez_robust_2012}.
Continuous dynamical decoupling (CDD) relies on the application of time-periodic continuous control fields, rather than a series of quantum-logic pulses.
Unlike conventional dynamical decoupling, CDD does not require any encoding overhead or quantum feedback measurements.

Thus far, CDD has been used to produce protected two-level systems in nitrogen vacancy centers in diamond, in nuclear magnetic resonance experiments, and trapped atomic ions~\cite{laucht_dressed_2017,farfurnik_experimental_2017,noguchi_generation_2012,golter_protecting_2014,timoney_quantum_2011,webster_simple_2013,barfuss_strong_2015,rohr_synchronizing_2014}, successfully inoculating them from spatiotemporal magnetic field fluctuations.
Here, we demonstrate CDD in an atomic Bose-Einstein condensate (BEC) producing a protected three-level system of dressed-states, whose Hamiltonian is fully controllable.
The CDD-protected states are sensitive to fluctuations of the amplitude of the control field itself, and we further demonstrate that a second coupling field protects against those in a concatenated manner~\cite{cohen_continuous_2017,farfurnik_experimental_2017,cai_robust_2012}.
\begin{figure}[t]
    \centering
    \includegraphics[]{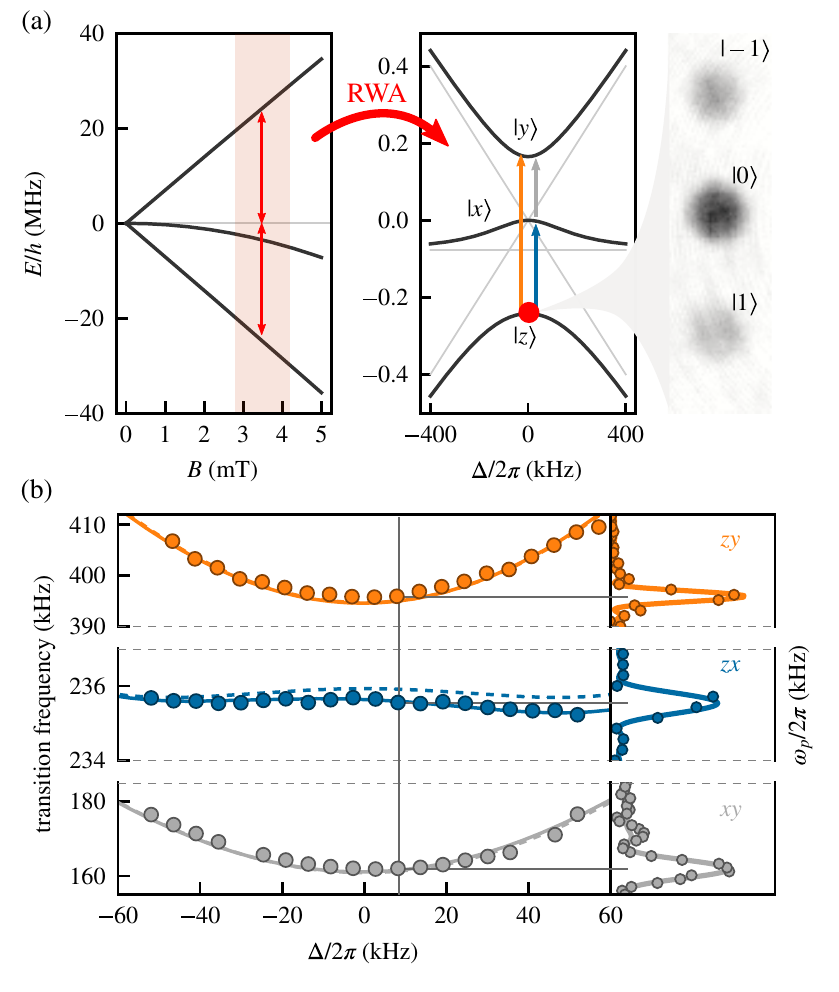}
    \caption{(a) Left: The dependence of the $5^2S_{1/2}$, $F=1$ ground state of \rb on an applied magnetic field, where the quadratic dependence of the $\ket{m_F=0}$ state's Zeeman shift has been exaggerated so it is visible on the same scale.
    Center: RWA eigenenergies of the \xyz eigenstates, evaluated for $\Omega/2\pi=\SI{200}{kHz}$ (black curves) and $\Omega=0$ (grey curves).
    Right: TOF image of $\ket{z}$ at $\Delta=0$, showing the decomposition into the constituent $m_F$ states.
    (b) Left: Spectroscopic data showing all possible transitions between the \xyz states for $\Omega/2\pi = \SI{194.5(1)}{kHz}$.
    Note that the vertical scale of the center panel marking the $zx$ transition has only 10\% the range of the other panels.
    Right: representative spectra.}
    \label{fig:1}
\end{figure}

\begin{figure}[t]
    \centering
    \includegraphics[]{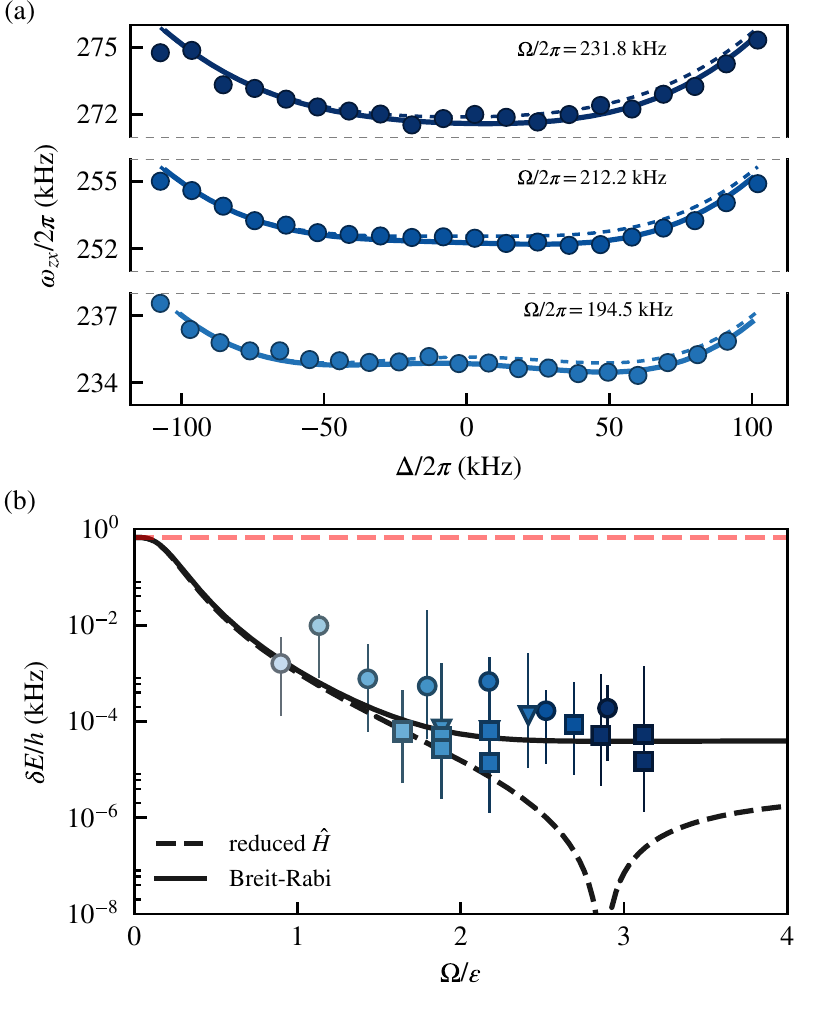}
    \caption{(a) Transition frequency $\omega_{zx}/2\pi$ measured for three different values of $\Omega/2\pi$, showing the minimal sensitivity to $\Delta$.
    The dashed curves correspond to \refeq{eq:h}, while the solid curves use the Breit-Rabi expression.
    (b) The change in energy associated with our typical experimental detuning fluctuations as measured in the $m_F$ basis is $\delta \Delta/2\pi = \SI{0.67}{kHz}$ (red dashed line).
    Triangles correspond to \xyz spectroscopy data, squares to side-of-peak $\pi$-pulse data, and circles to double-dressed data (see main text).
    The dashed curve is calculated using the Hamiltonian in \refeq{eq:h} and the solid curve using the Breit-Rabi expression.}
    \label{fig:2}
\end{figure}

\textit{System.}
We implemented CDD using a strong radio-frequency (RF) magnetic field of coupling strength $\Omega$, that linked the three $m_F$ states comprising the $F=1$ electronic ground state manifold of \rb.
The RF field was linearly polarized along ${\bf e}_x$, and had angular frequency $\omega$ close to the Larmor frequency $\omega_0 = g_F \mu_{\rm B} B_0$ from a magnetic field $B_0 {\bf e}_z$; $g_F$ is the Lande g-factor and $\mu_{\rm B}$ is the Bohr magneton.
We coupled the dressed states using a weaker probe field with coupling strength $\Omega_p$, polarized along ${\bf e}_y$ with angular frequency $\omega+\omega_p$ (\reffig{fig:1}a).
Using the rotating wave approximation (RWA) for the frame rotating at $\omega$ (valid when $\omega_0 \gg \Omega,\,\Omega_p,\,\omega_p$), the system is described by the Hamiltonian
\begin{equation}
    \hat H=\Delta\hat F_z + \Omega \hat F_x + \Omega_p \cos(\omega_p t) \hat F_y + \hbar\epsilon(\hat F_z^2 / \hbar^2 - \hat{\mathbb I}),
    \label{eq:h}
\end{equation}
where $\Delta=\omega-\omega_0$ is a detuning; $\epsilon$ is the quadratic Zeeman shift; $\hat F_{x,y,z}$ are the spin-1 angular momentum operators; and $\hat{\mathbb I}$ is the identity operator.
For $\Omega_p = 0$ the resulting eigenstates of \refeq{eq:h} are linear combinations of the $m_F$ states and we denote them as $\ket{x}$, $\ket{y}$, and $\ket{z}$.
The corresponding eigenvalues for $\Delta = 0$ are $\omega_x = 0$ and $\omega_{y,z} = -(\epsilon \pm \sqrt{4 \Omega^{2} + \epsilon^{2}})/2$.
The resulting energy differences $\hbar\omega_{xy}$, $\hbar\omega_{zy}$ and $\hbar\omega_{zx}$ are only quadratically sensitive to detuning $\Delta$ for $\Delta\ll\Omega$~\footnote{Although the energies scale quadratically with detuning for $\Delta\ll\Omega$, they scale linearly for $\Delta\gg\Omega$ with a slope of \SI{7}{MHz/mT}.} so that any fluctuations $\delta \Delta$ in the detuning are suppressed to first order, making these a trio of synthetic clock states.
At an optimal $\Omega$, $\omega_{zx}$ depends only quartically on $\Delta$~\cite{xu_coherence-protected_2012,rabl_strong_2009}.
For $\Delta \gg \Omega$ the \xyz states adiabatically connect to the corresponding $\ket{m_F=1,0,-1}$, states (see \reffig{fig:1}b).
As $\Omega\rightarrow 0$ and for $\Delta=0$, the \xyz states continuously approach the \XYZ states that are familiar from quantum chemistry where they form a common basis to represent atomic orbitals.
In contrast, as $\Omega\to\infty$ they become eigenstates of the $\hat F_x$ operator: $\ket{y,x,z} \to \ket{m_x=+1,0,-1}$.
Unlike for the $m_F$ basis, an oscillatory magnetic field can drive transitions between all pairs of the \xyz states with non-zero transition matrix elements.

For all the experiments described here, our BECs had approximately $N=\num{5e4}$ atoms, and were held in a crossed dipole trap with trapping frequencies $(f_x,\, f_y,\, f_z) = (42(3),\, 34(2),\, 133(3))$\,Hz~\footnote{All uncertainties herein represent the uncorrelated combination of statistical and systematic errors.}.
The $B_0 \approx \SI{3.27}{mT}$ bias field lifted the ground state degeneracy, giving an $\omega_0/2\pi = \SI{22.9}{MHz}$ Larmor frequency, with a quadratic shift $\epsilon/2\pi=\SI{76.4}{kHz}$.
In our laboratory the ambient magnetic field fluctuations were dominated by contributions from line noise giving an rms detuning uncertainty $\delta\Delta/2\pi = g_F \mu_{\rm B}\delta B/h=\SI{0.67(3)}{kHz}$.

We used an adiabatic rapid passage (ARP) technique to transfer atoms initially prepared in any of the $\ket{m_F = 0,-1,1}$ states into the corresponding \xyz states; this protocol began far from resonance at $\Delta(t=0)/2\pi \approx -\SI{450}{kHz}$ with all coupling fields off.
We then ramped on the RF dressing field in a two-step process.
First, we ramped from $\Omega=0$ to approximately half its final value in \SI{10}{ms}.
By increasing the magnetic field $B_0$, we then ramped $\Delta$ to zero in \SI{12}{ms} using a non-linear ramp chosen to be adiabatic with respect to the relevent energy gaps.
After allowing $B_0$ to stabilize for \SI{30}{ms}, we ramped the RF dressing field to its final value $\Omega$ in \SI{10}{ms}, yielding the dynamically decoupled system used in subsequent experiments.

To measure the population of the \xyz states, we adiabatically deloaded them back into the $m_F$ basis by ramping $B_0$ so that $\Delta$ approached its initial detuned value in \SI{2}{ms}, and then ramped off the dressing RF field in \SI{1}{ms}.
We obtained the spin-resolved momentum distribution using standard time-of-flight (TOF) imaging techniques, with an applied Stern-Gerlach field that spatially separated the different spin components during TOF.
The right panel of \reffig{fig:1}a shows the decomposition of $\ket z$ into the $m_F$ states in a typical TOF image.

We confirmed our control and measurement techniques by using the probe field to spectroscopically measure the energy differences between the \xyz states.
Figure~\ref{fig:1}b shows the dependence of the transition frequencies $\omega_{xy}/2\pi$, $\omega_{yz}/2\pi$, and $\omega_{zx}/2\pi$ on detuning for $\Omega/2\pi=\SI{194.5(1)}{kHz}$.
Each point corresponds to the peak location of a spectroscopical measurement at the given detuning $\Delta$; typical spectra for a probe with coupling strength $\Omega_p/2\pi \approx \SI{1}{kHz}$ and $\Delta/2\pi \approx \SI{9}{kHz}$ are shown on the side panel.
The dashed curves presenting the expected behavior based on \refeq{eq:h}, clearly depart from our measurements for the $zx$ transition.
This departure results from neglecting the weak dependence of the quadratic shift $\epsilon$ on bias field $B_0$.
In near-perfect agreement with experiment, the solid curves derived using the full Breit-Rabi expression account for this dependency.

\textit{Robustness.}
As shown in \reffig{fig:2}a, the $zx$ transition is remarkably robust against magnetic field variations, as commonly result from temporal and spatial magnetic field noise in laboratory environments
We now confine our focus to the $zx$ transition, which can be made virtually independent of magnetic field variations due to the similar curvature of $\omega_z(\Delta)$ and $\omega_x(\Delta)$ (see the middle panel of \reffig{fig:1}a).
We quantified the sensitivity of this transition to field variations using three different methods corresponding to the different markers in \reffig{fig:2}b.
In each of these cases we measured the energy shift from resonance as a function of detuning and then from a fourth order polynomial fit to the data computed the residual rms fluctuations $\delta \omega_{zx}$ due to magnitude of the known detuning noise~\footnote{Our fourth order procedure is able to quantify even the small fluctuations that survive for spectra that are flat through third order, such as our idealized model in \refeq{eq:h}.}.
Firstly, triangles denote data using full spectroscopical measurements similar to \reffig{fig:2}a.
Secondly, square markers denote data in which a detuned $\pi$-pulse of the probe field transferred atoms from $\ket z$ to $\ket x$, a side-of-peak technique giving a signal first-order sensitive to changes in $\omega_{zx}$.
Lastly, the circular markers describe data using an adiabatic technique described below.
The results for all the above methods agree fairly well with the theory using either \refeq{eq:h} (solid) or the Breit-Rabi expression (curved); both Hamiltonians give practically identical results on resonance~\footnote{The fluctuations can be even smaller for a given $\Omega$ when we are not constrained at $\Delta=0$ (see Supplemental Materials).}.

Even at the smallest value of $\Omega/2\pi=\SI{69(1)}{kHz}$ the typical magnetic field noise was attenuated by two orders of magnitude, rendering it undetectable.
Ideally, the radius of curvature of $\omega_{zx}(\Delta)$ changes sign at about $\Omega/2\pi = \SI{220}{kHz}$, leaving only a $\Delta^4$ contribution, however, in practice the small dependence of $\epsilon$ on $B$ prevents the cancellation leading to this flat point, and makes the residual linear term dominant instead.
\begin{figure}[t]
    \centering
    \includegraphics[]{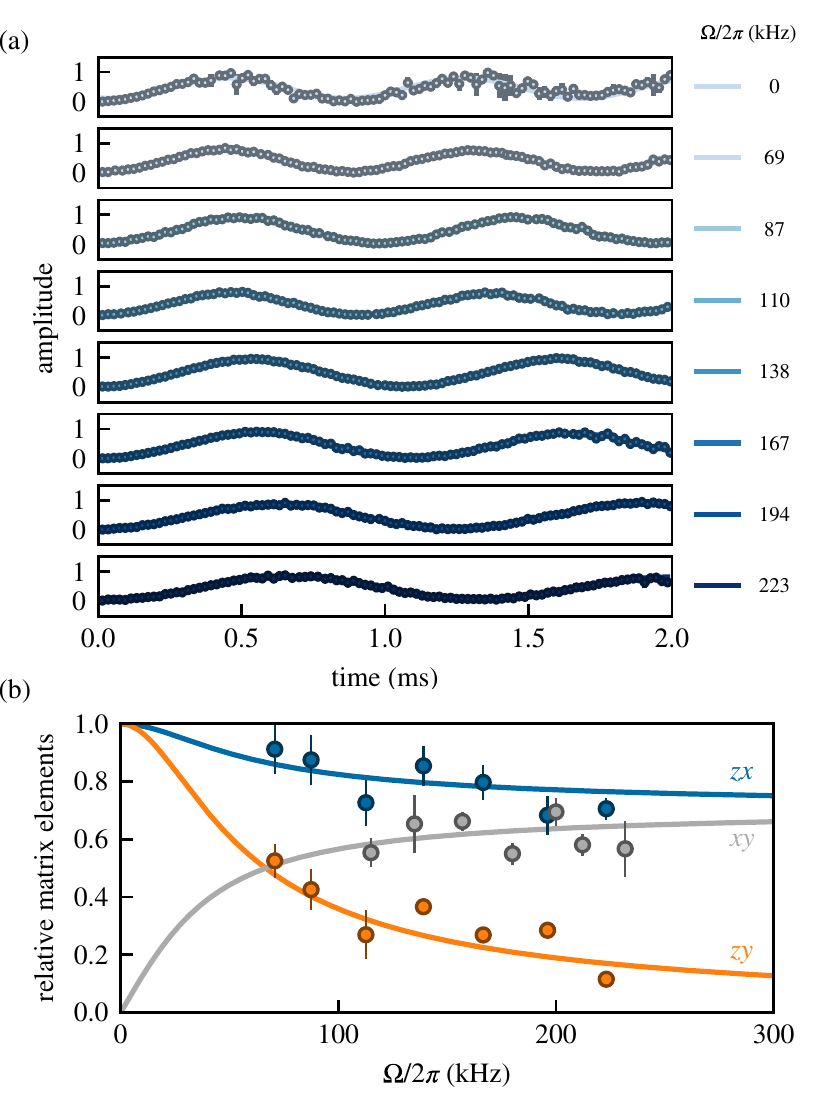}
    \caption{(a) Rabi oscillations for the various values of $\Omega$.
    Phase coherence is maintained throughout the oscillations in the dressed basis, while it is quickly lost in the $m_F$ basis.
    The marker size reflects the typical uncertainties on the dressed basis oscillations.
    (b) Transition matrix elements for $zx$ (blue) and $zy$ (orange) transitions decrease monotonically with increasing $\Omega$ for $\Delta=0$, while they increase for $xy$.
    This leads to an effective three-level system with only two allowed transitions (similar to the $m_F$ basis) for $\Omega \gg \epsilon$.
    }
    \label{fig:3}
\end{figure}

We explored the strength of the probe-driven transitions between these states by observing coherent Rabi oscillations, as shown in \reffig{fig:3}a.
With our BEC prepared in $\ket z$, the probe field was pulsed on at $\Omega_p/2\pi\approx\SI{1}{kHz}$.
The top panel shows Rabi oscillations between $\ket{m_F=0}$ and $\ket{m_F=-1}$ states for reference, and the remaining panels show oscillations between $\ket{z}$ and $\ket{x}$.
The observed Rabi frequency between dressed states decreased with increasing $\Omega$ indicating a dependence of the $zx$ transition matrix elements on coupling strength $\Omega$.
These matrix elements, as well as those for the $zy$ transition, decrease with increasing $\Omega$ for $\Delta=0$ as shown in \reffig{fig:3}b.
The coherence of the Rabi oscillations for longer times was limited by gradients in $\Omega$ that lead to phase separation of the dressed states, and therefore loss of contrast after a few tens of ms, but had no measurable effect on the coherence of the oscillations.
In comparison, the coherence of the Rabi oscillation between the $m_F$ states deteriorated significantly after \SI{500}{\us}.
For these timescales, the loss of coherence was predominantly due to bias magnetic field temporal noise~\footnote{We cancelled gradient magnetic fields so that no phase separation of the bare states was observed for $>\SI{10}{sec}$.}.

\textit{Concatenated CDD.}
The driving field $\Omega$ coupled together the $\ket{m_F}$ states, giving us synthetic clock states \xyz that were nearly insensitive to magnetic field fluctuations.
However, the spectrum of these states is first-order sensitive to the amplitude fluctuations $\delta \Omega$ of the driving field.
Reference~\cite{cai_robust_2012} showed that an additional field coupling together these \xyz states can produce doubly-dressed states that are insensitive to both $\delta \Omega$ and $\delta \Delta$: a process called concatenated CDD.
In our experiment, the probe field provided the concatenating coupling field.
Because $\Omega_p\ll\Omega$, we focus on a near-resonant two-level system formed by a single pair of dressed states, here $\ket{z}$ and $\ket{x}$, which we consider as pseudospins $\ket{\uparrow}$ and $\ket{\downarrow}$.
These are described by the effective two-level Hamiltonian
\begin{equation}
    \hat H_p = \frac{\hbar\Delta'}{2} \hat \sigma_3 + \hbar\Omega' \cos(\omega_p t) \hat \sigma_1,
    \label{eq:h2}
\end{equation}
with energy gap $\Delta' \approx \omega_{\downarrow, \uparrow}$ (shifted by off-resonant coupling to the $zy$ and $xy$ transitions) and coupling strength $\Omega' \propto \Omega_p$, as set by the matrix elements displayed in~\reffig{fig:3}b.
Here $\hat \sigma_{1,2,3}$ are the three Pauli operators.
\begin{figure}[t]
    \centering
    \includegraphics[]{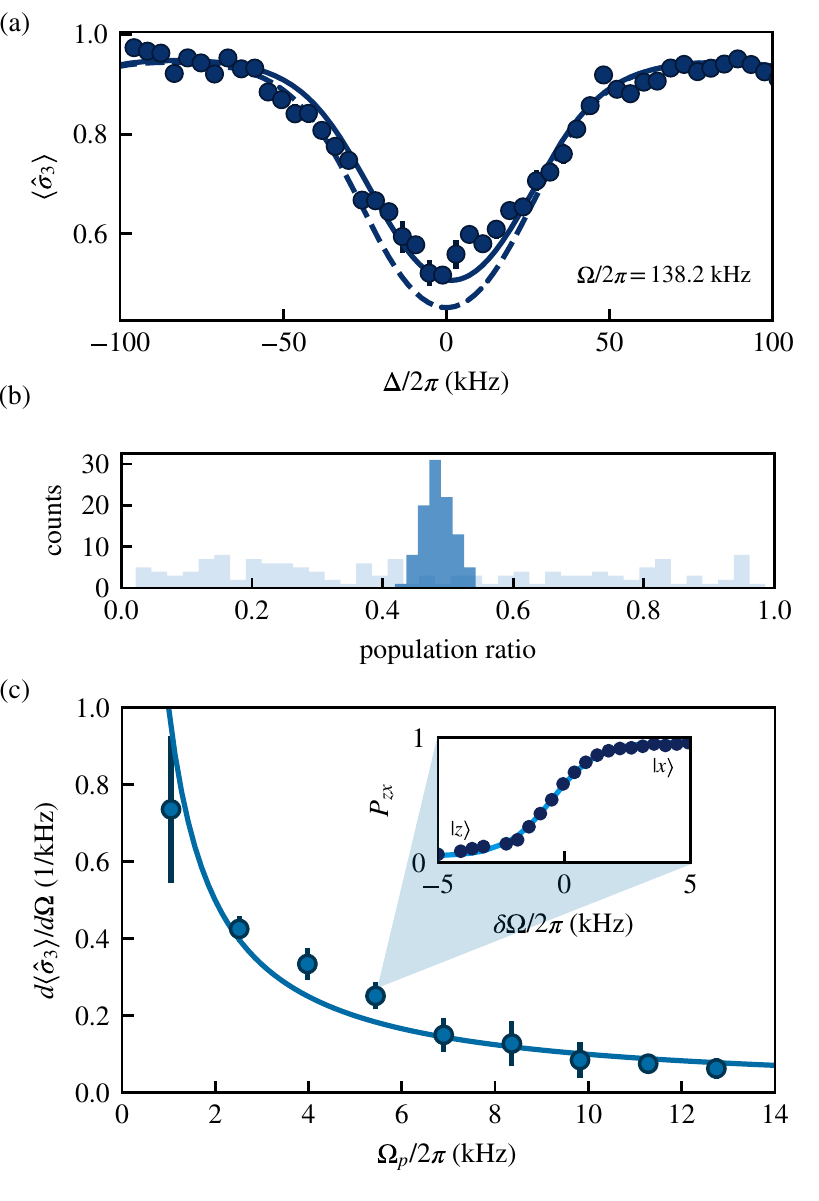}
    \caption{(a) The fractional population imbalance of the $\downarrow\uparrow$ transition for $\Omega/2\pi=\SI{138.2(1)}{kHz}$ over detuning $\Delta$.
    The dashed curve is calculated using \refeq{eq:h} and the solid one using the full Breit-Rabi expression.
    (b) The fidelity of preparing a balanced superposition of $\ket\downarrow$ and $\ket\uparrow$ (dark blue) states compared to $\ket{m_F=0}$ and $\ket{m_F=-1}$ states (light blue).
    (c) The robustness of $\downarrow, \uparrow$ transition against fluctuations $\delta \Omega$ for different probe field coupling strengths.
    The points represent the slope of the fitted curves to the fractional population imbalance (inset).}
    \label{fig:4}
\end{figure}
A RWA of this Hamiltonian leads to the energy spectrum $E_{\uparrow,\downarrow} \approx \pm\Omega^\prime/2 + (\Delta^\prime)^2/2\Omega^\prime$, having again assumed the coupling $\Omega^\prime$ exceeds any fluctuations in $\Delta^\prime$.
Thus, the concatenated CDD field protects from the fluctuations $\delta\Delta\prime$ of the first dressing field in the same way that CDD provided protection from detuning noise $\delta \Delta$.

We produced doubly-dressed states by using the probe field near resonant with the $\downarrow, \uparrow$ transition and an ARP sequence.
We started in $\ket\downarrow$ at $\Delta=0$ and ramped on the probe field $\Omega_p$ a few ms before ramping $\Omega$ to its final value.
We chose the ARP parameters such that we created an equal superposition of $\ket\downarrow$ and $\ket\uparrow$.
We quantified the sensitivity of this transition to large changes in the detuning by measuring the fractional population imbalance $\langle\hat\sigma_3\rangle = P_\downarrow(\Delta)-P_\uparrow(\Delta)$, shown in \reffig{fig:4}a for $\Omega/2\pi=\SI{138.2(1)}{kHz}$~\footnote{For large enough values of $\Delta$ the $zx$ and $xy$ transition become degenerate in energy and the system resembles an $F=1$ ground state at low magnetic field. We chose the maximum value of $\Delta$ such that the population of \unexpanded{$\ket y$}, which maps to \unexpanded{$\ket 1$} at large detuning, was negligible after deloading.}.
This signal is first-order sensitive to $\omega_{\downarrow, \uparrow}$, and provided our third measurement of sensitivity to detuning in \reffig{fig:2}b denoted by circles.

We compared the fidelity of preparing a superposition of the $\ket\downarrow$ and $\ket\uparrow$ states to adiabatically preparing a similar superposition of the the $\ket{m_F=0}$ and $\ket{m_F=-1}$ states.
The coupling strength of the probe field was about \SI{1}{kHz} in both cases.
Figure~\ref{fig:4}b shows the rms deviation of the population imbalance measured over a few hundred repetitions of the experiment.
The rms deviation for the dressed basis is $0.024(1)$ and is and order of magnitude smaller than for the $m_F$ basis $0.29(1)$, where it practically impossible to prepare a balanced superposition for the parameters used here~\footnote{In \reffig{fig:4}b, the noise in the $m_F$ basis in not Gaussian distributed as is typical of line noise in these experiments.}.

Figure~\ref{fig:4}c shows the response of the $\downarrow, \uparrow$ transition to small changes $\delta\Omega$ for different values of $\Omega_p$.
We prepared an equal superposition of $\ket\downarrow$ and $\ket\uparrow$ following the same procedure as before for $\Omega/2\pi = \SI{138.2(1)}{kHz}$.
We then measured how the population imbalance changes for small variations of $\Omega$ --- the effective detuning in the `twice-rotated frame' --- for different probe amplitudes $\Omega_p$.
We defined a sensitivity parameter $d\langle\hat\sigma_3\rangle / d\Omega$, obtained from the linear regime of the population imbalance measurements (see inset in \reffig{fig:4}c).
We found the robustness of the doubly-dressed states against $\delta \Omega$ fluctuations increased with $\Omega_p$, thus verifying the concatenating effect of CDD in the \xyz basis.

\textit{Conclusions.}
We realized and studied a three-level system that is dynamically decoupled from low-frequency noise and where direct transitions between all three states are allowed.
We demonstrated control techniques required to create arbitrary Hamiltonians in this three-level system; a feature that is not possible in the $m_F$ states.
These techniques add no heating or loss mechanisms, yet within the protected subspace retain the full complement of cold-atom coherent control tools such as optical lattices and Raman laser coupling, and permit new first-order transitions that are absent in the unprotected subspace.
These transitions enable experiments requiring a fully connected geometry as for engineering exotic states, e.g., in cold-atom topological insulators, and two-dimensional Rashba spin-orbit coupling in ultracold atomic systems~\cite{campbell_rashba_2016, juzeliunas_generalized_2010}.

The synthetic clock states form a decoherence-free subspace that can be used in quantum information tasks where conventional clock states might be absent, or incompatible with other technical requirements~\cite{bacon_universal_2000}.
Moreover, their energy differences are proportional to the amplitude of the dressing field, and hence tunable, so they can be brought to resonance with a separate quantum system.
The effective quantization axis can be arbitrarily rotated so that the two systems can be strongly coupled, pointing to applications in hybrid quantum systems~\cite{solano_chapter_2017,xiang_hybrid_2013}.
Introducing a second coupling field shields the system from fluctuations of the first, a process which can be concatenated as needed.
More broadly, synthetic clock states should prove generally useful in any situation where fluctuations of the coupling field can be made smaller than those of the environment.

\begin{acknowledgments}
This work was partially supported by the ARO’s atomtronics MURI, the AFOSR’s Quantum Matter MURI, NIST, and the NSF through the PFC at the JQI. We are grateful to P. Solano for carefully reading this manuscript.
\end{acknowledgments}

\bibliography{RFClockStates.bib}

%%%%%%%%%% Merge with supplemental materials %%%%%%%%%%
% \newpage
% \clearpage
\pagebreak
\widetext
\begin{center}
\textbf{\large Supplemental Materials}
\end{center}

\setcounter{equation}{0}
\setcounter{figure}{0}
\setcounter{table}{0}
\setcounter{page}{1}
\makeatletter
\renewcommand{\theequation}{S\arabic{equation}}
\renewcommand{\thefigure}{S\arabic{figure}}
\renewcommand{\bibnumfmt}[1]{[S#1]}
\renewcommand{\citenumfont}[1]{S#1}
% Prefix a "S" to all equations, figures, tables and reset the counter

\section{The \xyz dressed basis}
The reduced Hamiltonian of \refeq{eq:h} can be diagonalized analytically.
The eigenvalues for $\Delta=0$ are $\omega_x = 0$, and $\omega_{z,y} = - \epsilon \pm \tilde\Omega$, where $\tilde\Omega = \sqrt{4 \Omega^2 + \epsilon^2}$ is a generalized Rabi frequency.
The corresponding (non-normalized) eigenvectors are linear combinations of the $m_F$ basis states:
\begin{align}
    \ket x =& \ket{-1} + \ket{1}, \nonumber \\
    \ket y =& \ket{-1} -\frac{\epsilon + \tilde\Omega}{\sqrt 2 \Omega} \ket 0 + \ket 1, \\
    \ket z =& \ket{-1} -\frac{\epsilon - \tilde\Omega}{\sqrt 2 \Omega} \ket 0 + \ket 1. \nonumber
\end{align}

\begin{figure}[ht]
    \centering
    \includegraphics[]{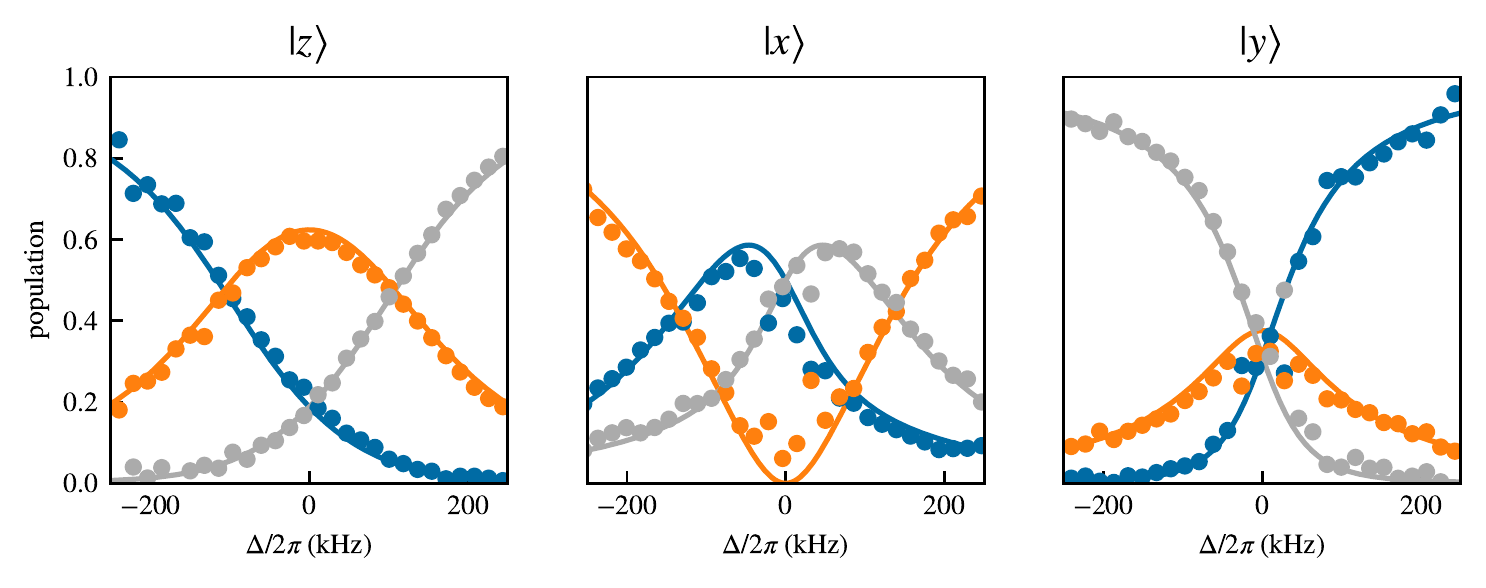}
    \caption{Decomposition of the \xyz states on the $m_F$ basis for $\Omega/2\pi=\SI{145(1)}{kHz}$
    The $\ket{m_F=-1,0,1}$ states correspond to blue, orange, gray respectively.}
    \label{fig:s1}
\end{figure}

We measured the above decomposition of the \xyz states to $m_F$ states using a projective measurement by abruptly turning off the dressing field $\Omega$ (see \reffig{fig:s1}).

For $\Delta=0$ and small coupling $\Omega / \epsilon \to 0$ with regard to the quadratic shift the $\ket y$ and $\ket x$ become symmetric and antisymmetric superpositions of the $\ket{m_F=-1, 1}$ states while $\ket z$ is predominantly composed of $\ket 0$
\begin{equation}
    \ket x = \ket{1} - \ket{-1}, \quad
    \ket y = \ket{1} + \ket{-1} + \frac{\Omega}{\epsilon}\ket 0, \quad
    \ket z = \frac{\Omega}{\epsilon}(\ket{1} + \ket{-1}) -\ket 0.
\end{equation}
On the other hand, when $\Omega\to\infty$ they are independent of the driving field amplitude and continuously approach the eigenstates of the $\hat F_x$ operator
\begin{equation}
    \ket x = \ket{1} - \ket{-1}, \quad
    \ket y = \ket{1} + \sqrt 2 \ket 0 + \ket{-1}, \quad
    \ket y = \ket{1} - \sqrt 2 \ket 0 + \ket{-1}.
\end{equation}
The states adiabatically map to the $\ket{m_F}$ states for $\Delta \gg \Omega$ as shown in \reffig{fig:s1}.
For $\Delta/2\pi > \SI{200}{kHz}$ the \xyz states are not yet fully deloaded to a single $m_F$ state since the population in other $m_F$ states is not negligible.
$\ket z$ maps to $\ket 1$ ($\ket{-1}$) for positive (negative) detuning; $\ket y$ maps in the exact opposite way to $\ket z$; and $\ket x$ always maps to $\ket 0$.

\subsection{Dependence on detuning}
For nonzero values of the detuning $\Delta$, the eigenvalues are the root of the characteristic cubic polynomial $H(\lambda)=\Delta^2\epsilon + (\Delta^2 + \Omega^2) \lambda - \epsilon \lambda^2 - \lambda^3$.
The eigenvalues are even functions with respect to $\Delta$ as can be seen by the leading order expansion for $\Delta\to 0$
\begin{align}
    \omega_x =& -\frac{\epsilon}{\Omega^2} \Delta^2 + \mathcal{O}(\Delta^4), \nonumber \\
    \omega_y =& \frac 12 (-\epsilon + \tilde\Omega) - \frac{(\epsilon + \tilde\Omega)}{-\epsilon^2-4\Omega^2+\epsilon\tilde\Omega} \Delta^2 + \mathcal{O}(\Delta^4), \label{eq:exp} \\
    \omega_z =& \frac 12 (-\epsilon - \tilde\Omega) + \frac{(\epsilon - \tilde\Omega)}{\epsilon^2+4\Omega^2+\epsilon\tilde\Omega} \Delta^2 + \mathcal{O}(\Delta^4). \nonumber
\end{align}

hence their resemblance to clock states.
We focused on the $zx$ transition since the curvature of $\omega_x$ and $\omega_z$ has the same sign for $\epsilon < \tilde \Omega$ (\refeq{eq:exp}).
Since the quadratic term changes curvature it can be made arbitrarily small.
However, this cancellation does not take place when we consider the dependence of $\epsilon$ on $\Delta$ from the Breit-Rabi expression.

\subsection{Transition matrix elements}
The \xyz states transform under the application of the spin-1 operators as $\epsilon_{jkl}\hat F_j \ket k= i\hbar \ket l$, so that a resonant probe field can induce transitions between at least one pair of states, irrespectively of its polarization.
\begin{figure}[ht]
    \centering
    \includegraphics[]{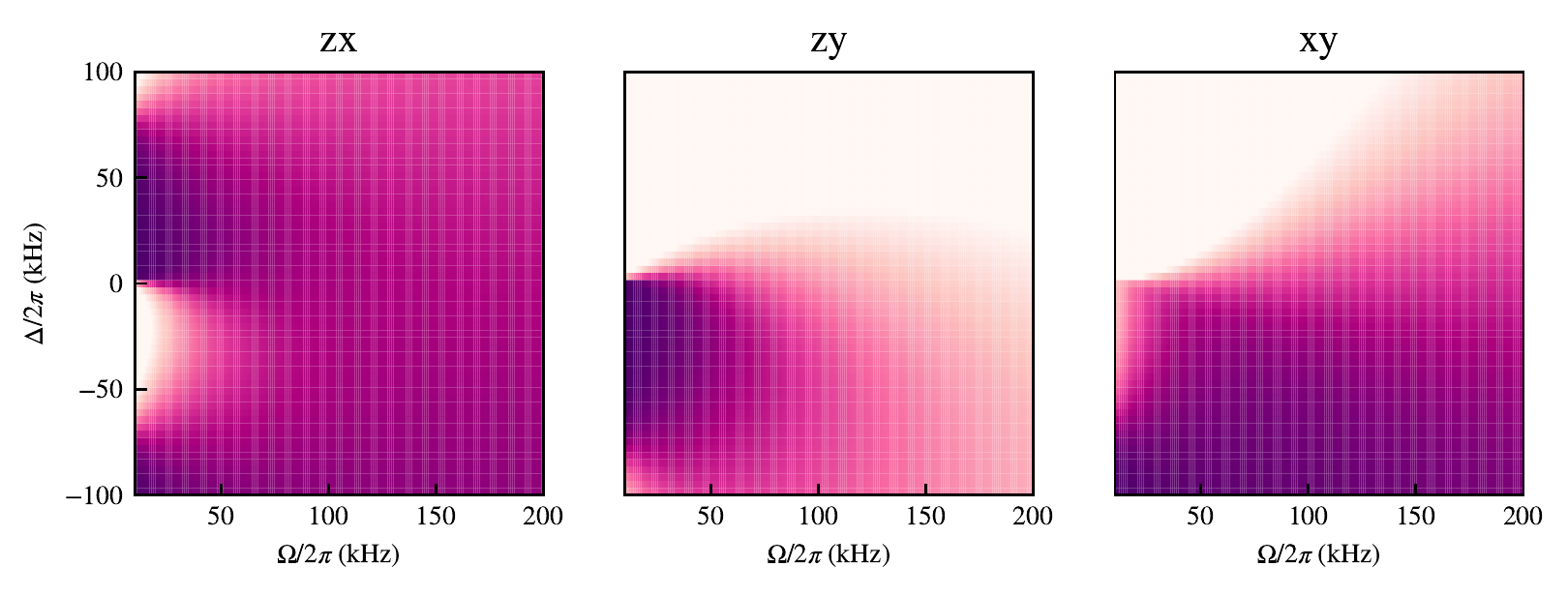}
    \caption{Transition matrix elements over $\Omega$ and $\Delta$.
    There is an assymetry between coupling on the blue and red side of the resonance that corresponds to the counter- and co-rotating terms $\hat F_-$ and $\hat F_+$.}
    \label{fig:s12}
\end{figure}

The transition matrix elements between the \xyz show a dependence on both $\Omega$ and $\Delta$ (see \reffig{fig:s12}).
For $\Omega \ll \epsilon$ the matrix elements correspond to those of the $\ket{m_F}$ basis and $\langle x \vert \hat F_+ \vert y \rangle = 0$ as expected by angular momentum selection rules.
When $\Omega$ and $\epsilon$ are comparable in magnitude al transition matrix elements are nonzero and the states can be coupled cyclically.
As $\Omega \gg \epsilon$ the $\ket z$ and $\ket y$ states decouple and the system resembles an `undressed basis' following similar selection rules.

\subsection{Optimal response to noise}

\begin{figure}[ht]
    \centering
    \includegraphics[]{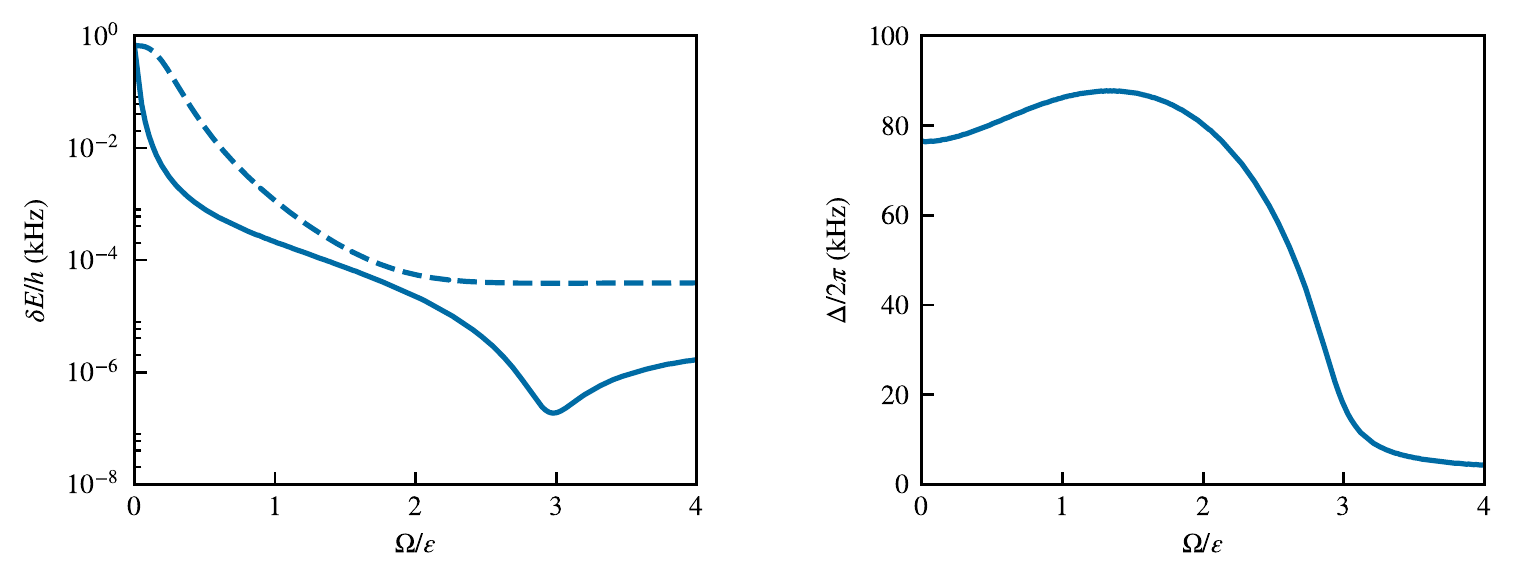}
    \caption{Left: The optimum response (solid) of the $zx$ transition to detuning fluctuations allowing for finite $\Delta$ compared to $\Delta=0$ (dashed).
    Right: The values of $\Delta$ that correspond to the minimum derivative of $\omega_{zx}$.}
    \label{fig:sopt}
\end{figure}

The sensitivity of the $zx$ transition to detuning fluctuations can be optimized further by working at $\Delta \neq 0$ as shown in \reffig{fig:sopt}.
This behavior can only be captured by including the dependence of the quadratic shift on $\Delta$ as given by the Breit-Rabi expression.

For small values of $\Omega$ the optimum value of $\Delta$ corresponds to on of the concave features of the $zx$ transition energy that arise due to the asymmetry introduced by the quadratic shift.
As $\Omega$ gets larger, these features merge into a single one and the optimum value is $\Delta \approx 0$.
The deviation from $\Delta=0$ is due to an overall tilt of the transition energy coming from the dependence of the quadratic shift on $\Delta$.
At the optimum point $\Omega/\epsilon \approx 3$ the sensitivity of the synthetic clock transition is \SI{1.9e-07}{kHz}, c.f, the \rb clock transition which scales as \SI{57.5}{kHz/mT^2} and gives \SI{5.8e-07}{kHz}.

\section{Locating field resonance}

\begin{figure}[ht]
    \centering
    \includegraphics[]{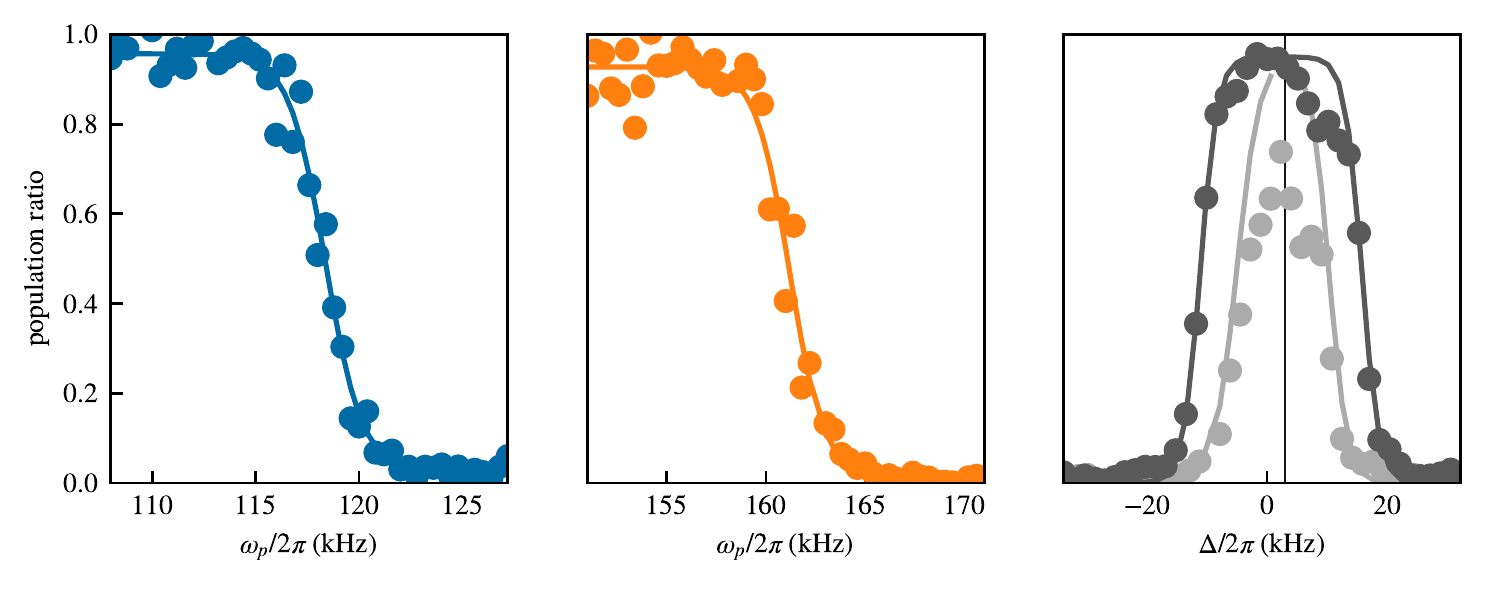}
    \caption{Characteristic spectroscopy curves for the $zx$ (left) and $zy$ (middle) transitions.
    Two symmetric ARPs (right) define the resonant value of the magnetic field.
    The width of the peak gets smaller as $\Delta$ gets closer to the resonant value.
    Here $\Delta/2\pi=\SI{3}{kHz}$.}
    \label{fig:s2}
\end{figure}

We used an iterative procedure to measure and adjust the value of the detuning $\Delta$ to account for the weak response of the \xyz states to detuning variations.
As most of our experiments were done at $\Delta=0$, we first obtained an estimate of $\Delta$ from the imbalance of $\ket 1$ and $\ket{-1}$ populations from the decomposition of $\ket z$ which should be zero for $\Delta=0$ (see \reffig{fig:s1}).
We then located the transition frequencies for at least two transitions (usually $zx$ and $zy$ as shown in \reffig{fig:s2}) using an ARP protocol as described in the main text and varying the frequency of the probe field.
These frequencies correspond to a unique pair of $\Omega$ and $\vert\Delta\vert$ values which can then be used to adjust the bias magnetic field $B_0$ so that $\Delta=0$.
However, there is an ambiguity as to the sign of $\Delta$ since the eigenstates are even functions of $\Delta$.

We selected a direction randomly and subsequently verified if $\Delta=0$ using another set of spectroscopic measurement.
We fixed the value of the probe field to be a few kHz above the transition frequency corresponding to $\Delta=0$ and used the same ARP sequence to transfer atoms from $\ket z$ to $\ket x$.
This procedure gave two resonant values for $\Delta$ were atom transfer takes place, and the value where $\Delta=0$ corresponded to their mean (see \reffig{fig:s2}).
Finally, we remeasured the $zx$ and $zy$ transition frequencies to validate that $\Delta=0$.
For higher values of $\Omega_R$, using the $zx$ transition becomes impractical
due to its insensitivity to detuning, and we followed the same procedure but using the $xy$ transition instead.

\end{document}